\documentclass[10pt,conference]{IEEEtran}

\usepackage{graphicx}
\usepackage{url}
\usepackage{cite}
\usepackage{hyperref}
\usepackage[formats]{listings}
\usepackage{tcolorbox}
\usepackage{adjustbox}
\usepackage{colortbl}

\lstdefineformat{R}{~=\( \sim \)}
\begin{document}

\title{Exit the Code: A Model for Understanding Career Abandonment Intention Among Software Developers}


\author{\IEEEauthorblockN{Tiago Massoni}
\IEEEauthorblockA{
Federal University of Campina Grande\\
Email: massoni@dsc.ufcg.edu.br}
\and
\IEEEauthorblockN{Ricardo Duarte}
\IEEEauthorblockA{
Paraiba State University\\
Email: ricardo.estat@yahoo.com.br}
\and
\IEEEauthorblockN{Ruan Oliveira}
\IEEEauthorblockA{
Brazilian Hospital Services Company\\
Email: ruan.oliveira@ebserh.gov.br
}}

\maketitle

\begin{abstract}
    \textit{Background.} Career abandonment -- the process in which professionals leave the activity, assuming positions in another area -- among software developers involves frustration with the lost investment and emotional and financial costs, even though being beneficial for the human being, depending on personal context.
    Previous studies have identified work-related motivators for career abandonment, such as the threat of obsolescence, unstable requirements, and low code quality, though these factors have primarily been examined in former developers. 
    The relationship between these motivators and the intention to abandon among currently active developers remains unexplored.
    \textit{Goal.} This article investigates the relationship between key work-related motivators and currently active software developers' intention to abandon their careers.
    \textit{Method.} We employed a quantitative approach, surveying 221 software developers to validate a theoretical model for career abandonment intention, based on an adaptation of the Investment Model, which incorporates satisfaction with technical aspects of the profession as well as the intention to abandon.
    \textit{Findings.} Exploratory and confirmatory factor analyses, through structural equation modeling (SEM), provided robust support for the adapted Investment Model in explaining software developers' intention to abandon their careers. 
    Moreover, career commitment significantly impacts the intention to leave the profession, being positively influenced by satisfaction with technical work-related factors and negatively influenced by career alternatives and career investment.
    \textit{Conclusion.} The paper offers valuable insights for organizational leaders and research, potentially guiding retention strategies to better support developers, and the adoption of theoretical models to explain career abandonment.
\end{abstract}


\section{Introduction}

One anonymous user published, in 2024, the following post in a forum for software developers on Reddit (the \href{https://www.reddit.com/r/conselhodecarreira/comments/1cu6hzp/carreira_de_12_anos_em_ti_bem_sucedida_s%C3%B3_consigo/?share_id=q3rWovdCf9wt7vDC6nhJC&utm_content=1&utm_medium=ios_app&utm_name=ioscss&utm_source=share&utm_term=1}{original post} in Portuguese):

\begin{quote}
\emph{(\ldots) The thing is, now at 34 years old, I don't see any more prospects. (\ldots) I'll spend another 25 years doing the exact same thing, (\ldots) it hurts to see younger people excited about things I've already done 50 times and getting recognition while I'm falling behind.
(\ldots) It feels like I'm wasting my time.}
\end{quote}

There has been relatively limited focus on a phenomenon known as \textit{career abandonment} \cite{scholtz-2019-role,colomo2014career} (or \textit{turnaway}), while \textit{developer turnover} has been extensively investigated, both for the IT workforce \cite{moore2000one,McKnight-2009-Which} and software developers\cite{ma2020data,bao2017will,turnover-2022,turnover-pandemic-2024}. 
Abandonment entails professionals exiting the software development field to assume different roles outside software development or entirely different occupations~\cite{joseph2015turnover}.

One must consider the \textit{social cost} of career abandonment; even though we understand career transition could be beneficial for the human being, depending on their context, such a drastic decision involves \textit{frustration with the lost investment} made throughout years of study and dedication, besides the \emph{emotional and financial costs} required for the professional change (Section~\ref{background}).
This cost encompasses not only the individual's frustration and potential financial losses but also the \emph{broader impact on the software industry}, especially given the high demand for skilled developers.

The decision to abandon a software development career is driven by various motivators, prominently \emph{psychological and emotional}, \emph{career-related}, and \emph{work-related}~\cite{colomo2014career}.
The latter --- including factors (motivators) \textit{intrinsic to the software development activity}, concerning dissatisfaction with core development tasks, such as coding, debugging, requirement elicitation or documentation, or even the threat of professional obsolescence  --- has not received as much attention as the other categories~\cite{colomo2014career,fu-2015-career,scholtz-2019-role}.

In this study, we survey 221 currently active software developers working in Brazil. 
The survey instrument combines items related to work-related motivators observed in the literature with the Investment Model~\cite{rusbult1983} -- explained in Section~\ref{theory} -- to explain their relationship with the \textit{intention to abandon the career} (Section~\ref{setup-quant}).
According to this theoretical model, individuals weigh their investments in a profession (e.g., time, effort, skill development) against the perceived rewards and available alternatives.
In this context, dissatisfaction with core software development tasks or the constant need for skill updates may amplify the sense that the return on investment is insufficient, thereby driving the decision to abandon the career.
The primary research question guiding this work can be stated as:
\emph{Does the investment model, adapted to software development, explain the intention to abandon a career?}

The resulting theoretical model was statistically analyzed by both Exploratory and Confirmatory Factor Analyses, the latter using Structural Equation Modeling (SEM) (Section~\ref{results-quant}).
The findings confirm the \emph{applicability} of the adapted investment model to \emph{explain the intention to abandon} a software development career, 
highlighting the significant role of satisfaction with core software development activities.
The results show that career commitment significantly impacts the intention to leave the profession, 
being positively influenced by satisfaction with technical work-related factors and negatively influenced by career alternatives and career investment.
We believe this research brings valuable insights for organizations seeking to develop and implement effective retention strategies for software developers, while potentially employed by researchers to improve theoretical models for the phenomenon (Section~\ref{discussion}).

\section{Career Abandonment}
\label{background}

Theories about \textit{Career Change} often focus on the interplay of individual, organizational, and environmental factors that drive transitions, 
given the different situations professionals face throughout their lives \cite{brown2013career}. 
Transitions can be motivated both by the personal decision of the worker who wants to seek new challenges and by other situations that impose the need to adapt to new realities, such as the loss of a job or a proposal to change work.
According to Hall \cite{hall-1996-protean}, careers in the 21st century will be ``protean'' -- the professional's career can be reinvented from time to time by themselves as the person or the environment changes.
This is especially sound for software developers -- members of a professional software development team.

Considering all the challenges and requirements of the modern and competitive world, some careers tend to suffer more with the transition process, such as IT and software development professionals. 
Technological advances and their wide use make them constantly required, having highly technical capacity, flexibility, and mobility, which can cause difficulty in staying motivated and responding to intense challenges \cite{loogma-2004-identification}.
In this context, two issues are often observed with these professionals: \textit{turnover} --- they switch jobs but remain in software development --- and \textit{abandonment, or turnaway} - they leave the profession, assuming a position in another functional area of the company or outside it~\cite{bellini-2019-should}. 
Often professional satisfaction and the need for growth are factors that can make a professional want to change jobs or profession \cite{brooks-2015-identifying,scholtz-2019-role}. 

Retaining professionals has been a challenge faced by most companies in recent years \cite{brooks-2015-identifying,tallon-2011-competing}. 
In this sense, it is essential to understand the expectations of these professionals about work and career and what aspects contribute to their permanence in the IT area or transition to other areas.
Understanding the career transition process that can culminate in career abandonment is a continuous challenge for researchers and professionals \cite{joseph-2007-turnover}.
In addition to professional development, Colombo-Palacios \cite{colomo2014career} identified characteristics such as financial motives, frustration, physical and psychological exhaustion, and lack of productivity as factors driving transition and career abandonment.
Other factors can influence the intention to abandon IT or software development: age, work experience, change complexity, obsolescence, professional self-efficacy, job insecurity, conflicts between family life and work, lack of autonomy, and unpleasant work environment \cite{scholtz-2019-role,ruan-sbes}.

\section{Theory Development}
\label{theory}

\subsection{Investment Model}

Several theories can inform the experience of career transitions, such as the \emph{Social Cognitive Career theory (SCCT)}, or \emph{Life-span/life-space Career Development Theory} \cite{brown2013career}.
Potential factors influencing career transitions stem from individual factors, such as skills, adaptability, or personal circumstances, and environmental factors, such as economic conditions and market forces.

The results of our previous exploratory study highlighted the importance of a low career investment and a high number of alternatives outside software development for the decision to abandon the profession \cite{ruan-sbes}.
As such, the Investment Model \cite{investment2011} provides a framework for predicting the state of being committed to someone or something, and for understanding the underlying causes of commitment, extending some basic principles of interdependence theory \cite{interdependence1980}.
Despite its focus on commitment to relationships, the model's principles are often adapted and applied to various contexts, including \emph{workplace}. 

In the context of job satisfaction and commitment, the model suggests that an employee's commitment to their job is influenced by three key factors: \textit{satisfaction, alternatives, and investment}.
According to the model, higher levels of job satisfaction --- contentment an employee experiences within their current job role --- lead to stronger commitment, due to the \emph{positive correlation between contentment and dedication} to the job.
Also, \textit{with attractive alternatives, employees might be less committed} to their current job, as they believe they can easily transition elsewhere. 
In this model, \textit{investment} may take various forms, including time, effort, skills developed, and emotional attachments. 
In the context of IT workers, Fu and others have explored the Investment Model \cite{fu-2011-understanding,fu-2015-career} to indicate antecedents of career commitment, naming expelling forces driving people away from their profession \textit{push effects}, closely related to our concept of career abandonment.



\subsection{Work-related Motivators to Career Abandonment}

Previous research has investigated which factors are linked to career abandonment in the IT and software development sectors.
For instance, our recent work explored tasks, processes, events, or situations that could have been a motivator for the decision to abandon the software development career~\cite{ruan-sbes,ruan-thesis}. 
Semi-structured interviews were conducted with 25 individuals who participated, at some point in their careers, in software teams as developers, analysts, testers, architects, or managers, but have transitioned to a diverse work area. 
After qualitative analysis, we came out with several work-related (technical) motivators; we show the most cited ones in Table~\ref{tab:mot}.

\begin{table}[ht]
    \caption{Technical Motivators for abandoning software development~\cite{ruan-thesis}.}
    \label{tab:mot}
    \centering
    \begin{tabular}{ll}
    \hline
    \textbf{\#Cit.}         & \textbf{Motivator}                                                                                    \\ \hline
    
    \multicolumn{1}{r}{39} & \multicolumn{1}{l}{Poor task planning}                                                          \\ \hline
    \multicolumn{1}{r}{33} & \multicolumn{1}{l}{Poor requirements}                                                    \\ \hline
    \multicolumn{1}{r}{28} & \multicolumn{1}{l}{Rework; Poor estimation}                                                                                           \\ \hline
    \multicolumn{1}{r}{23} & \multicolumn{1}{l}{\begin{tabular}[c]{@{}l@{}}Ever-changing requirements\end{tabular}}                                                                      \\ \hline
    \multicolumn{1}{r}{16} & \multicolumn{1}{l}{Threat of professional obsolescence}                                                                                         \\ \hline
    \multicolumn{1}{r}{11} & \multicolumn{1}{l}{Lack of documentation; Low code quality}                                                                                         \\ \hline
\end{tabular}  
\end{table}

\textit{Poor task planning} stands out as the most frequently cited issue, indicating that ineffective task management is a significant source of frustration for developers, likely leading to feelings of inefficiency and burnout. 
The second most cited motivator is \textit{poor requirements}, along with \emph{ever-changing requirements}, reflecting how critical clear and well-defined requirements are in maintaining motivation and engagement in software projects.
In addition, \emph{excessive rework} suggests that issues with project planning and effort estimation are recurring challenges in software development. 
Concerns on \emph{low code quality} and \emph{lack of documentation} are also relevantly mentioned as motivators for abandonment.
Regarding \emph{threat of professional obsolescence}, unlike other professions, however, in which knowledge and skills dissolve at a less accelerated pace, the lifespan of the technical domain for a software professional is perceived as much shorter, so developers are obliged to be constantly updating and iterating about new technologies, programming language, and frameworks for development \cite{colomo2014career}. 

These technical motivators inspired us to build categories of work-related constructs, including them as the \emph{Satisfaction} component of the Investment Model -- see Section \ref{sec:adapting} for more details on the adaptation.

\section{Methodology}
\label{setup-quant}

We aim to apply the Investment Model to elucidate the phenomenon of career abandonment in software developers through a survey. 
Given the challenges in recruiting a sample of former software developers, our focus lies on studying actively employed developers, utilizing a scale of intention to abandon the career. This methodology aligns with previous approaches in career-transition investigations within the IT and software development domains \cite{colomo2014career,joseph-2007-turnover,bao2017will}.
The ethics for this study were reviewed and approved by the appropriate Institutional Review Board\footnote{\url{https://plataformabrasil.saude.gov.br/}, reference number\:29946520\-.2.\-0000.\-5182}. 

In this study, \textit{we directed our focus towards technical work-related aspects} of software development. 
To integrate the Investment Model with the most cited motivators, we incorporate them into the survey instrument, within the \textit{Satisfaction with Software Development} (SAT\_SD) component. 
Subsequently, we applied statistical factor analysis to assess this adapted model, measuring the relationship of its factors to abandonment intention.
Furthermore, the survey data and statistical scripts are available as an online replication kit~\cite{rep-kit}.

\subsection{Study Research Questions}

\textbf{Q1.} \textit{What factors emerge from the survey instrument assessing software-specific motivators and career abandonment intention?} We employ Exploratory Factor Analysis (EFA) to evaluate the validity of the survey instrument, in particular the new satisfaction items, ensuring that the observed variables align with the theoretical constructs of the adapted Investment Model.

\textbf{Q2.} \textit{How do the components of the Investment Model --- specifically, satisfaction with software development, career investment, alternatives, and commitment --- collectively explain the intention to abandon among software developers?} Utilizing Confirmatory Factor Analysis (CFA) and structural equation modeling, we assess the fit of our adapted Investment Model to elucidate the relationships between latent variables and the intention to abandon.

This survey-based quantitative research aims to identify a population or sample characterized by specific factors or capable of justifying a particular event. 
Quantitative research is employed to evaluate respondents' explicit and conscious opinions and attitudes, utilizing standardized instruments.
Our underlying philosophical assumption aligns with \textit{postpositivism} \cite{creswell2017book}, acknowledging \textit{the existence of an objective reality that can be observed and measured}. 
While embracing postpositivist principles, which recognize the importance of empirical observation and data collection in forming hypotheses, testing theories, and drawing tentative conclusions, the study also acknowledges inherent limitations in our understanding.

\subsection{Adapting the Investment Model}
\label{sec:adapting}

\begin{figure}[htbp]	
    \centering
    \includegraphics[width=1\linewidth]{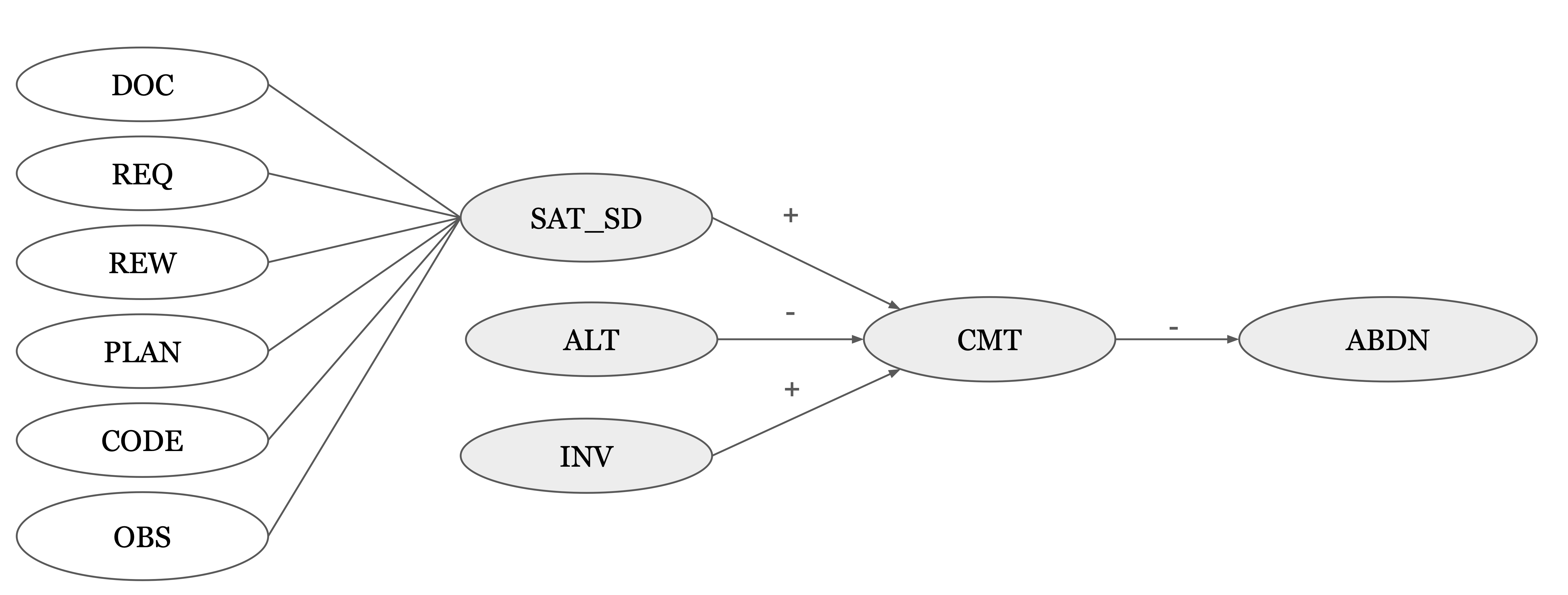}
	\caption{Adapted version of the Investment Model.}
	\label{fig:adapted-models}
\end{figure}

To develop the survey instrument, we categorized motivators found in our previous study \cite{ruan-sbes,ruan-thesis} into six main groups: \textit{Poor Documentation (DOC), Poor Requirements (REQ), Excessive Rework (REW), Poor Planning (PLAN), Low Code quality (CODE), and Threat of professional obsolescence (OBS)}. 
For example, various motivators related to requirements, such as Ever-changing requirements, Lack of good requirements, and Requirements elicitation, were consolidated into a group labeled REQ.
This categorization was conducted by one author and subsequently reviewed by another, with any disparities discussed and resolved. 
Subsequently, we chose to utilize these groups as the \textit{Satisfaction (SAT\_SD)} factor within the Investment Model.
Additionally, we incorporated other constructs from the theory, including availability of career alternatives (ALT) and career investment (INV). 
The main dependent variable, abandonment intention (ABDN), was employed as a surrogate for the original job stability variable. 
Figure \ref{fig:adapted-models} depicts the adapted model, with Career Commitment (CMT) as a mediator, which assesses the potential mediating role of \textit{Commitment} in the causal pathways between our independent and dependent variables. 

The survey questionnaire, structured with 38 items rated on a five-point Likert scale, incorporated established measures and existing instruments to ensure the reliability of the study and enable comparisons with related research.
To assess the intention to abandon the career, the Meyer and Allen model, a well-recognized instrument for measuring job turnover intention, was utilized, comprising four items \cite{jaros1997assessment}.
The availability of career alternatives was evaluated through a three-item scale adapted from Dam \cite{van-2005-employee}, alongside a six-item career investment scale. 
The Threat of professional obsolescence (OBS), adapted from Pazy and Kaufman \cite{meyer-1993-commitment}, consisted of four items in this study. Two items assessed the scope, evaluating required knowledge, while the other two measured how quickly professionals realized their knowledge had become obsolete.
CMT was evaluated using the nine-item scale developed by Carson and Bedeian \cite{kidd-2006-careers}. However, only four items were employed in this study, with three related to resilience and one to career identification.
For the software-inherent motivators, a new set of items for each group was developed, inspired by the Perceived Workload instrument by Kirmeyer and Dougherty \cite{kirmeyer-1988-work}. 
The complete survey instrument is available in Appendix~\ref{append}.


Following the translation of the items into Portuguese, a pilot study was conducted to validate the instrument. 
The pilot involved 20 voluntary software developers, contributing to the clarification and identification of ambiguity in the items' texts. 
Subsequently, a reliability and validity test of the constructs comprising the questionnaire was performed using the pilot data; 
the pilot was not used in the final data set.
In the instrument design, \textit{item inversion} was employed to enhance reliability and internal consistency, particularly for ALT, CMT, and ABDN.
By incorporating inverted items, the aim was to mitigate the risk of participants falling into response patterns compromising validity.
In the final data, scores for those items were corrected accordingly.
By calculating the Cronbach's Alpha for the final sample, 
we observed that all values surpass $0.8$, 
a widely accepted threshold indicating a \textit{good} level of reliability \cite{taber2018cronbach}.

\subsection{Procedure and Sample}

The research population consists of currently active software developers working in Brazil. 
The sample was defined using \emph{non-probabilistic convenience sampling}~\cite{baltes2021sampling}, 
chosen to foster a more substantial number of responses.
To increase the number of respondents and enhance the diversity of the sample, the following strategy was employed: 
an e-mail containing information about the research and a link to the questionnaire was sent to \emph{hubs} -- 
individuals in leadership positions within Brazilian software organizations who could distribute the survey instrument to a broader group of software developers.

In the final phase, efforts were focused on including additional participants from underrepresented sociodemographic groups, 
with a particular emphasis on women and specific regions of Brazil. 
To achieve this, new hubs and interest groups relevant to these categories were identified. 
For example, an NGO dedicated to promoting women in technology was contacted to assist in further disseminating the instrument.
The questionnaire was administered using the online tool Survey Monkey \footnote{\url{https://www.surveymonkey.com/}}, resulting in a \emph{valid dataset comprising 221 software developers over three months between September and December 2022}.
Due to the influential individuals' distribution of the survey link, the response rate could not be calculated. 

After accumulating a significant number of responses 
(the sample size of 221 aligns with prior studies on career transition in the IT industry \cite{colomo2014career,fu-2015-career}), 
the data preparation for analysis commenced, 
discarding the partially completed answers.
The data were formatted in an electronic spreadsheet and imported into RStudio Desktop \footnote{\url{https://posit.co/download/rstudio-desktop/}}.

\subsection{Analysis Method}

Due to the proposed modification of the Investment Model, \emph{Exploratory Factor Analysis} (EFA) is employed to verify whether these hypothesized factors align with latent variables \cite{thompson2004exploratory}.
EFA serves to reduce the dimensionality of the observed variables, ensuring that the identified factors are elucidated by the primary observed components. 
The initial step involves constructing a \textit{Correlation Matrix} based on Spearman's correlation \cite{zar2005spearman} to examine variables with a high mutual correlation. 
After the value surpasses the threshold of $0.50$, indicating a factorable matrix, the Kaiser-Meyer-Olkin (KMO) criterion and the Bartlett Sphericity Test ascertain the validity of EFA for the selected variables.
Parallel Analysis is then used as the factor extraction method, followed by Oblimin oblique rotation with normalization; with this, EFA can be effectively applied to data that do not meet strict normality assumptions. \cite{watkins_step_by_step_2020}.
Finally, attention is given to the \textit{eigenvalue}, a crucial indicator in EFA that signifies the number of latent factors explaining the items \cite{thompson2004exploratory}.

In executing \emph{Confirmatory Factorial Analysis} (CFA), a \emph{Structural Equation Model} (SEM) is applied \cite{harrington2009confirmatory}. 
This technique serves to evaluate the goodness of fit of a given theoretical model to the issue under investigation, along with the correlational structure emerging from the data.
To gauge model fit, the RMSR (Residual Mean Square Root) measures the discrepancy between observed and predicted values. Lower RMSR values indicate a better fit, with a value approaching zero suggesting an excellent fit. 
Further, the RMSEA (Root Mean Squares of Approximation Errors) accounts for model complexity by penalizing for overfitting. An ideal RMSEA value is below $0.05$, signifying a well-fitting model.
Finally, the TLI (Tucker-Lewis Index) measures the relative improvement of the model over a null model. 
A TLI close to $1$ indicates a good fit, with values above $0.9$ generally considered acceptable.




\section{Results}
\label{results-quant}



Most participants were men ($67\%$) --- $28\%$ women, $5\%$ not informed. The age distribution has a prevalence of developers between 20 and 40 years old ($41\%$). 
Regarding Brazil's geographical regions, over 77\% of respondents are located in the three most economically developed areas (Northeast, South, and Southeast). 
The typical participant ($59.7\%$) holds a bachelor's degree and identifies themselves as programmers ($49.8\%$) or quality assurance professionals ($23.5\%$). 
Smaller organizations, with up to 250 employees, predominate ($37\%$).

\subsection{Q1: What factors emerge from the survey instrument assessing software-specific motivators and career abandonment intention?}

Table \ref{tab:correlations} shows a correlation map for the items of the survey instrument.   
For all observed variables, we amalgamated strongly correlated items into a composite score, computed as \textit{the sum of individual scores}, to improve visualization. 
For instance, items REQ1, REQ2, REQ3 and REQ4 are added up to produce REQ.
Asterisks mean highly significant correlations ($p<0.001$), and no asterisks mean the correlation is not statistically significant, suggesting it may have occurred by chance.

\begin{table*}[ht]
    \caption{Correlation matrix for the composite survey items.}
    \label{tab:correlations}
    \centering
    \begin{tabular}{lrrrrrrrrrr}
    \hline
                  & \multicolumn{1}{c}{\textbf{DOC}} & \multicolumn{1}{c}{\textbf{REQ}} & \multicolumn{1}{c}{\textbf{REW}} & \multicolumn{1}{c}{\textbf{PLAN}} & \multicolumn{1}{c}{\textbf{CODE}} & \multicolumn{1}{c}{\textbf{OBS}} & \multicolumn{1}{c}{\textbf{ALT}} & \multicolumn{1}{c}{\textbf{INV}} & \multicolumn{1}{c}{\textbf{CMT}} & \multicolumn{1}{l}{\textbf{ABDN}} \\ \hline
    \rowcolor[gray]{0.9} 
    \textbf{DOC}  &                                  & 0.85*                            & 0.83*                            & 0.87*                             & 0.74*                             & - 0.81*                          & - 0.61*                          & 0.05                             & 0.50*                            & - 0.69*                           \\
    \textbf{REQ}  &                                  &                                  & 0.92*                            & 0.90*                             & 0.80*                             & - 0.88*                          & - 0.70*                          & 0.03                             & 0.60*                            & - 0.78*                           \\
    \rowcolor[gray]{0.9} 
    \textbf{REW}  &                                  &                                  &                                  & 0.90*                             & 0.81*                             & - 0.89*                          & - 0.73*                          & 0.03                             & 0.64*                            & - 0.81*                           \\
    \textbf{PLAN} &                                  &                                  &                                  &                                   & 0.75*                             & - 0.89*                          & -0.68*                           & 0.01                             & 0.54*                            & - 0.75*                           \\
    \rowcolor[gray]{0.9} 
    \textbf{CODE} &                                  &                                  &                                  &                                   &                                   & - 0.79*                          & - 0.70*                          & 0.16*                            & 0.61*                            & - 0.80*                           \\
    \textbf{OBS}  &                                  &                                  &                                  &                                   &                                   &                                  & 0.76*                            & - 0.06                           & - 0.65*                          & 0.87*                             \\
    \rowcolor[gray]{0.9} 
    \textbf{ALT}  &                                  &                                  &                                  &                                   &                                   &                                  &                                  & -0.28*                          & - 0.59*                          & 0.83*                             \\
    \textbf{INV}  &                                  &                                  &                                  &                                   &                                   &                                  &                                  &                                  & - 0.01                            & - 0.27*                           \\
    \rowcolor[gray]{0.9} 
    \textbf{CMT}  &                                  &                                  &                                  &                                   &                                   &                                  &                                  &                                  &                                  & - 0.70*                           \\ \hline
    \end{tabular}
    \end{table*}

The higher DOC, REQ, REW, PLAN, and CODE values, the higher the satisfaction with that specific activity, while lower values for OBS represent the participant feeling more threatened by obsolescence. Higher ALT and INV represent, respectively, more perceived career alternatives and career investment. The participant committed to the career should answer with higher CMT scores. Finally, higher ABDN scores represent a higher intention to abandon the career.

Before proceeding to EFA, the basic assumptions for factor analysis must be met. 
The correlation matrix showed a KMO $=0.97$, representing high factorability \cite{stevens-2012-applied}. The Bartlett's test results ($\chi^2=11423.27,df=703,p < 0.001$) suggest strong evidence to reject the null hypothesis that the variances are equal; \emph{the data is considered then appropriate for EFA}.

Table \ref{tab:result-eigenvalues} shows the \textit{eigenvalues}, the variance explained by each factor, and the accumulated variance for each variable.  
The eigenvalues represent the sum of the column of squared factor loadings for a factor, in which the result is suggested to be greater than one (eigenvalue $>1$). 
The score we measured is approximately $1$ as of Factor 4. 
As a result, four factors were able to explain approximately $80\%$ of the total variance, a percentage considered good, despite all predictor variables being maintained.
We applied \textit{Oblique Orthogonal Rotation} due to the likely correlation among the factors, with \textit{Ordinary Least Squares} as factoring method; this method provides satisfactory results of maximum likelihood without assuming a multivariate normal distribution \cite{varimax1994}.
We label and describe the four factors as follows.

\begin{table}[ht]
\caption{Results of eigenvalues.}
\begin{center}
 \begin{tabular}{|p{45mm}|p{45mm}|p{45mm}|p{45mm}|}
\hline
\multicolumn{1}{c}{\textbf{ID.Factor}}  & 
\multicolumn{1}{c}{\textbf{Eigenvalues}}  & \multicolumn{1}{c}{\textbf{Var.porc}} & \multicolumn{1}{c}{\textbf{Var.acum}}
\\ \hline
\multicolumn{1}{l}{F1}  & \multicolumn{1}{l}{23.77} & \multicolumn{1}{l}{62.55} & \multicolumn{1}{l}{62.55} 
\\ \hline         

\multicolumn{1}{l}{F2}  & \multicolumn{1}{l}{3.37} & \multicolumn{1}{l}{8.87} & \multicolumn{1}{l}{71.42} 
\\ \hline

\multicolumn{1}{l}{F3}  & \multicolumn{1}{l}{2.3} & \multicolumn{1}{l}{6.06} & \multicolumn{1}{l}{77.48} 

\\ \hline   

\multicolumn{1}{l}{F4}  & \multicolumn{1}{l}{0.98} & \multicolumn{1}{l}{2.58} & \multicolumn{1}{l}{80.06} 

\\ \hline   

\end{tabular}  
\end{center}
\label{tab:result-eigenvalues}
\end{table}

\noindent{\textbf{F1 -- Satisfaction with Software Development (DOC, REQ, REW, PLAN, CODE, OBS).}} Details of the first factor are related to the overall satisfaction with the software development-related tasks previously identified as motivators for abandoning the career in our qualitative study\cite{ruan-thesis}. 

\noindent{\textbf{F2 -- Career investment (INV).}} The variables for INV represent a second concept, which seems orthogonal to the other career aspects. 

\noindent{\textbf{F3 -- Career commitment (CMT).}} This third factor includes scores that reflect how involved the professionals are with their current career status. 

\noindent{\textbf{F4 -- Abandonment (ALT,ABDN).}} From the loadings, we assume a stronger association between the presence of career alternatives and the intention to abandon software development.


\begin{tcolorbox}[colback = white, left=0pt, right=0pt, top=0pt, bottom=0pt] \label{Q1}
    \textbf{Answer to Q1}: Latent variables F1 (DOC, REQ, REW, PLAN, CODE, OBS), F2 (INV), F3 (CMT) and F4 (ALT, ABDN) appropriately represent the key dimensions from the Investment Model. Also, the observed relationships confirm that \emph{the items accurately reflect the theoretical constructs they were designed to measure}.
\end{tcolorbox}

\subsection{Q2: How do the components of the Investment Model --- specifically, satisfaction with software development, career investment, alternatives, and commitment --- collectively explain the intention to abandon among software developers?}

While the EFA allows us to observe a clear separation of the observed variables into latent variables, CFA provides a more in-depth view of the relationship between these elements, in order to evaluate how the investment models fit into the career abandonment intention by developers. 

\subsubsection{Model Fitness}

Before Structural Equation Modeling (SEM), recommended indices were calculated to evaluate the consistency of the factorial structure. 
Table \ref{tab:cfa-ind} presents each index, the values obtained from the data sample, and the reference values used as benchmarks. 
Notably, the RMSEA was slightly higher than the proposed limit ($0.08$), although still close to the marginal threshold suggested by the literature \cite{stevens-2012-applied}.
CFA results indicate that each observed variable loaded significantly onto its respective latent variable, affirming the measurement structure. 
The model-fit indices provide strong evidence that the selected model aligns well with the survey data, thereby offering \emph{strong support for applying the adapted investment model to the analysis of career abandonment intention}.

\begin{table}[htpb]
\caption{Model-Fit Indices for Confirmatory Factor Analysis (CFA)}
\begin{center}
\begin{tabular}{lll}
\hline
\multicolumn{1}{l}{\textbf{Index}}             & \multicolumn{1}{c}{\textbf{Value}} & \multicolumn{1}{c}{\textbf{Reference Value}} \\ \hline
TLI                         & 0.903                               & TLI$\geq$0.9                                  \\ \hline
CFI                       & 0.91                                & CFI$\geq$0.9                                  \\ \hline
RMSEA   & 0.085                               & RMSEA$\leq$0.08                               \\ \hline
\end{tabular}
\end{center}
\label{tab:cfa-ind}
\end{table}

\subsubsection{Path Analysis}

\begin{figure*}[t]	
    \centering
    \includegraphics[width=0.75\linewidth]{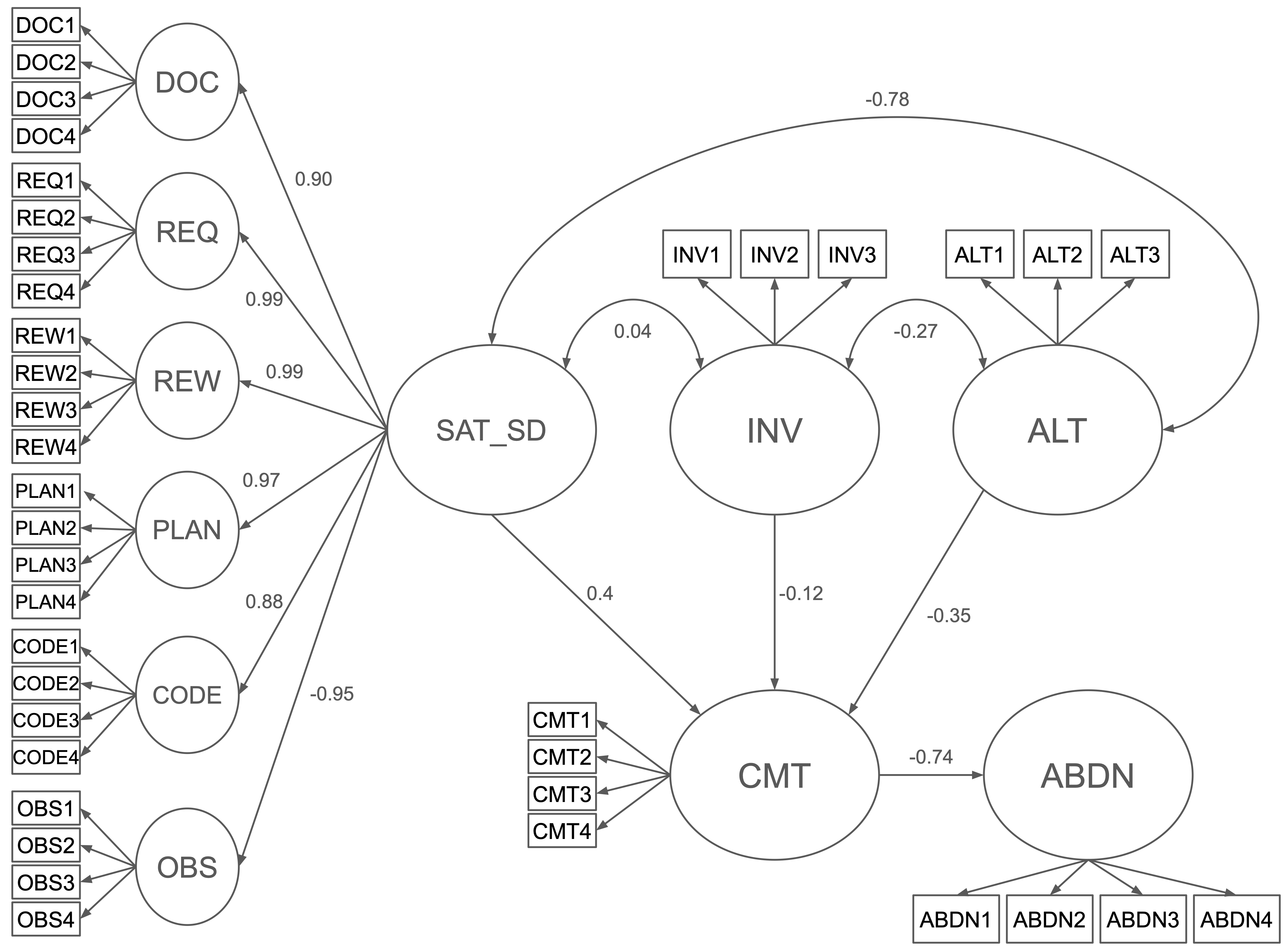}
	\caption{Path Diagram with Loadings.}
	\label{fig:path-diagram}
\end{figure*}

Once the measurement models are validated, we can proceed with path models (structural models) that link the latent variables, as shown in Figure~\ref{fig:path-diagram}. 
The standard notation represents latent variables (SAT\_SD, INV, ALT, CMT, and ABDN) as ellipses, 
and their corresponding observed variables (e.g., INV1, DOC2) as rectangles. 
SEM model loadings are illustrated with path coefficients as arrows between variables. 
Rounded edges depict covariances between latent variables, though several covariance paths are omitted for clarity. 
The loadings are labeled with their \textit{standardized path coefficient}, 
estimating the strength of the regression paths; 
all shown coefficients are \emph{statistically significant}.

The variable \textit{Intention to Abandon} is primarily explained by the mediating variable \textit{Career Commitment}, 
with a strong negative regression path of $-0.74$, emphasizing the substantial impact of career commitment on lower intention to abandon the career. 
In turn, commitment is strongly supported by \textit{Satisfaction with Software Development} ($0.4$) and \textit{Career Alternatives} ($-0.35$), 
suggesting that greater satisfaction with the software development field and fewer career alternatives are associated with higher commitment levels. 
One could consider the negative support of \textit{Career Investment} ($-0.12$) to commitment counterintuitive. 
It may show a misalignment of investment and personal aspirations that lead to commitment, especially if individuals feel that their investments are not yielding expected rewards.
Overall, \textit{Career Commitment} as a mediator clarifies that the relationship between satisfaction, investment, alternatives, and career abandonment is indirect, channeled through the degree of an individual's commitment to their career.

\begin{tcolorbox}[colback = white, left=0pt, right=0pt, top=0pt, bottom=0pt] \label{q2}
    \textbf{Answer to Q2}: The adapted Investment Model \emph{effectively captures the factors influencing individuals' intention to abandon their careers}. Career commitment has a significant impact on the intention to leave a career, and this commitment is positively influenced by Satisfaction with Software Development, while it is negatively influenced by Career Alternatives and Career Investment. 
\end{tcolorbox}

\section{Discussion}
\label{discussion}

\subsection{Discussion of the results}

The correlation matrix (Table~\ref{tab:correlations}) reveals robust correlation clusters among software-inherent satisfaction factors, notably \emph{positive} correlations for DOC, REQ, REW, PLAN, CODE, and a \emph{negative} correlation for OBS, an expected result from the chosen theoretical model. 
The expected relationship among the first five factors was confirmed by our data --- a developer unsatisfied with one of them tends 
to dislike the other issues. 
Interestingly, individuals feeling threatened of obsolescence tend to lack satisfaction, for IT workers in general, making them consider career change~\cite{obsolescence-IT}.
A positive correlation of $0.76$ exists between the presence of career alternatives and the threat of obsolescence, 
indicating \emph{an individual may be searching for other career paths while being affected by the feeling they are staying behind technically}.

Furthermore, the moderate negative correlations with career alternatives imply an inverse relationship with satisfaction variables. 
This suggests that \emph{dissatisfaction with certain aspects of software development prompts individuals to explore alternative career paths}, 
as observed in the literature~\cite{fu-2011-understanding}.
Career investment does not significantly correlate with other variables, except the presence of career alternatives ---  
in this case, a moderate negative correlation, as observed in previous research \cite{fu-2015-career}, 
which suggests a mild conflict between how much the individuals invested and the number of career options available to them.
The correlation between ABDN and CMT is $-0.65$, indicating that higher levels of abandonment intention are negatively 
associated with commitment, which has also been reported before \cite{jaros1997assessment,fu-2015-career,fu-2011-understanding}.

Regarding the satisfaction with software activities, from EFA's resulting factors, 
the data suggests that professionals present similar feelings around issues with documentation, requirement management, rework needs, code quality and task planning. 
Professionals showing higher satisfaction with these aspects also tend to be less concerned with obsolescence, 
at the same time having fewer career alternatives available. 
This factor aggregates scores related to the technical side of the profession, as selected for this study.
Investment and commitment, as expected, appear as orthogonal dimensions to those technical aspects, 
while more career alternatives are highly associated with a higher intention to abandon. 
This is expected, as an individual, possessing ways to transition to another occupation 
(e.g. family business and enough savings to start a new business), 
\emph{understandably would have more reasons to express a will to abandon software development}.

From the results of CFA, 
the strong negative support of CMT for ABDN indicates that 
career commitment is markedly associated with a lower intention to leave the profession. 
It is expected that committed professionals 
--- showing passion for the work, a sense of professional identity, or a belief in how they contribute to the team ---
are less likely to consider another type of work~\cite{rusbult1983}. 
Fostering a strong sense of commitment among developers is certainly a way to maintain them as active professionals, 
although commitment depends on other factors, such as job satisfaction, a supportive work environment and recognition\cite{fu-2015-career}. 
In this sense, our results partially corroborate with conclusions from previous research~\cite{fu-2011-understanding,colomo2014career}.

As commitment is positively supported by SAT\_SD,
indirectly affecting the intention to leave the profession, 
it is clear that enhancing satisfaction with the activities is key for long-term retention.
The issue is that an individual being satisfied with the technical aspects of their profession is not solely due to the work environment; 
other factors certainly play a role, such as an individual's aptitude for the profession.
On the other hand, negative support from ALT and INV is related to commitment inversely. 
The presence of appealing alternative career paths can undermine commitment, as expected. 
On the other hand, a small negative relationship between INV and CMT may seem counterintuitive 
but can be interpreted in several ways depending on the nuances of the data. 
If individuals feel that their investments are not yielding the expected rewards or personal growth, they may become less committed to their careers.
Also, high levels of investment might result in burnout or exhaustion, which can diminish an individual's commitment.
Future qualitative studies will be able to clarify context-specific relationships.

The positive covariance of $0.04$ between SAT\_SD and INV suggests a weak positive relationship, 
indicating that individuals who report higher levels of satisfaction may also tend to invest more in their career 
--- the inverse also holds. 
Conversely, the negative covariance of $-0.27$ between INV and ALT indicates a moderate negative relationship, 
indicating that as individuals invest more in their careers, they perceive fewer viable alternatives. 
Similarly, the substantial negative covariance of $-0.78$ between SAT\_SD and ALT suggests that individuals reporting higher levels of satisfaction with software development perceive fewer viable alternatives.

A good fit of our adaptation of the investment model indicates that the model's pathways 
--- such as the effects of career investment, available alternatives, and, in particular, satisfaction with key software-related activities --- 
adequately capture the factors influencing individuals' intention to abandon their careers.
As the original Investment Model has been used to explain career commitment in other professions~\cite{investment2011}, 
this result is expected. 
Still, our conceptualization of satisfaction resonates well with the collected data, 
indicating that \emph{these variables are indeed a critical component in understanding the intention to abandon a career}.

\subsection{Implications}

\noindent\textbf{For Researchers.} The findings, particularly the strong fit of the updated model with the adjusted dimension (SAT\_SD), suggest that the Investment Model can be adapted to specific careers, such as software development. This presents opportunities for future research to further refine and expand the model across various professional contexts. Additionally, the significant role of career commitment as a mediator among satisfaction, alternatives, investment, and career abandonment underscores the need for researchers to explore other potential mediators and their effects. Such exploration could lead to the development of alternative models that explain why individuals choose to remain in or leave their careers.
Our research employs technical factors as proxies for the satisfaction dimension in the Investment Model. However, as highlighted in previous studies \cite{colomo2014career,ruan-sbes,scholtz-2019-role}, emotional and psychological factors (such as stress and work-life balance), along with elements of the work environment, could significantly influence career abandonment. Integrating these components into a cohesive model could represent a valuable direction for future investigation.

\noindent\textbf{For Practitioners.} The strong relationship between factors specific to the profession (such as poor requirements and the constant need for rework) and commitment to the career suggests that intrinsic enjoyment and fulfillment from daily work tasks are critical for retaining software developers in their roles.
Lack of satisfaction appears to have a stronger influence on career commitment than a lack of investment or the availability of numerous career alternatives.
Based on these results, it is imperative that developers reflect on whether their dissatisfaction with certain tasks arises from the inherent nature of the task itself or from incidental factors.
For example, unstable requirements might stem from the inherent complexity of the domain or from poor requirement management within the organization.
In the latter case, developers could actively engage with team members and leaders to address these issues or consider seeking a position in a different organization that better aligns with their preferences.
Additionally, they might evaluate whether transitioning to a different role within the software development field (e.g., management, UX design, or research) is a viable option.

Nevertheless, some people might benefit from leaving software development. 
Individuals may find that transitioning to different roles provides a better work-life balance, a reduced workload, and lower stress levels, all of which can enhance mental and emotional well-being \cite{ruan-sbes}.
Furthermore, exploring alternative career paths can present new opportunities for personal and professional growth, 
especially when viable options are available 
(as indicated by the negative indirect effect of alternatives on abandonment). 

\noindent\textbf{For Team Leaders and Managers.} Since our results suggest career commitment significantly influences the decision to leave a profession, leaders should be vigilant for signs that a developer's commitment is waning. 
Retention strategies could involve providing incentives to retain key team members, preventing burnout, and creating opportunities 
for both personal and professional development to make alternative career options less appealing. 
Understanding the unique strengths and preferences of each team member -- and assigning roles accordingly -- is crucial for enhancing job satisfaction, which is the most significant factor linked to career commitment.
For this, leaders should prioritize improving communication with developers, 
ensuring that there are clear channels for understanding their challenges, aspirations, and frustrations. 
Furthermore, leaders should recognize instances where a developer's long-term happiness might be better served by exploring different career opportunities. In such cases, providing support and incentives for transitioning away from software development may be the most considerate approach. By demonstrating empathy and genuine concern for the overall well-being of developers, leaders can foster a culture of trust, which positively impacts team morale and loyalty over time.

\subsection{Limitations}

\noindent\textbf{Construct validity.}  
Abandonment is considered a complex, multi-factorial, and poorly understood phenomenon in software development, considering most studies trying to find factors correlating with abandonment intention \cite{bellini-2019-should}. 
Our research focuses on understanding how work-related factors, particularly satisfaction with core software development activities, interact to either minimize or exacerbate the phenomenon of career abandonment. 
The results are not conclusive, as it primarily centers on work-related variables. 
Other significant factors, such as career-related variables (e.g., long-term growth opportunities, job security) and psychological-emotional variables (e.g., burnout, emotional exhaustion), may also play a crucial role in influencing the decision to abandon a software development career \cite{ruan-sbes,colomo2014career}. 
Besides, being a sample study, causal relationships
is typically impossible; our hypotheses rather propose associations between different constructs.

\noindent\textbf{External validity.} The survey was conducted online and anonymously, so we are unable to provide details about how representative the sample was. 
It was a convenience sample, gathered through professional networks and social media. 
However, since our study targeted software professionals, it is unlikely that individuals outside the software industry would have participated. 
The sample includes a diverse range of software professionals, which supports the aim of our study—to gather evidence for our theoretical model focused on this group.
That said, additional studies are needed to replicate these findings.
Also, the study was conducted in a single country (Brazil), thus regional and cultural differences were not accounted for; each region has unique factors that may influence development contexts.

\noindent\textbf{Construct validity.} We modified established measurement tools and created derived instruments for some constructs based on previous research. 
Our assessment of the measurement model indicated that our constructs were internally reliable and performed well on tests of both convergent and discriminant validity. 
Additionally, we replaced the original Satisfaction component from the Investment Model with a composite construct, as we did not find a similar construct in the existing literature that reflects the same concept. 
Despite being newly defined, this construct performed strongly in the validity tests and EFA.

\noindent\textbf{Reliability and Objectivity.} 
Participants' responses might be influenced by their current mood or recall biases, affecting consistency over time.
Even though we did not collect responses in different times and contexts, the survey focuses on recent events and experiences, asking participants to report on specific, short-term activities rather than relying on distant memories, in order to reduce recall bias. 
Additionally, providing clear and detailed prompts in the survey might have helped participants accurately remember and reflect on past experiences, improving the reliability of their responses.
Also, the study employed convenience sampling and relied on influential individuals to distribute the survey, potentially introducing biases in participant selection and affecting the generalizability of findings. 
We applied established scales and instruments to promote reliability and incorporated item inversion to mitigate social desirability bias.

\section{Related Work}
\label{related}

Our adaptation of the Investment Model is based on our previous qualitative study~\cite{ruan-sbes,ruan-thesis}, 
which identified two types of motivators for career abandonment in software developers: 
general and software development-specific. 
The latter inspired us to develop the survey instrument, 
using the most frequently cited motivators as proxies for the Satisfaction construct in our model.
Another relevant study applied a mixed-methods approach to identify career abandonment factors, beginning with a qualitative analysis of 20 software developers' intentions to leave, followed by a quantitative analysis of 148 responses to measure correlations \cite{colomo2014career}. 
This study identified 22 factors associated with career abandonment in software, with perceived workload, threat of obsolescence, instability, lack of autonomy, and absence of a formal career path having the strongest correlations. While this study provided a broad overview, our research specifically adapts a theoretical model to this phenomenon, highlighting factors like professional obsolescence, career investment, commitment, and career alternatives.

Fu examined the push–pull–mooring (PPM) framework alongside the Investment Model to explore IT professionals' career retention and abandonment \cite{fu-2011-understanding}.
Push effects, such as low satisfaction and professional obsolescence, drive individuals away, while pull effects, such as appealing career alternatives, attract them toward new paths.
Mooring variables, such as career investment and self-efficacy, anchor individuals to their profession.
While both studies utilize the Investment Model, Fu's work integrates it with the PPM framework and examines a broad range of IT professionals, focusing on general career-related factors such as professional self-efficacy.
In contrast, our study applies a modified version of the Investment Model, emphasizing technical, work-related factors specific to software developers.
Similarly to what Fu found for IT professionals, our results suggest that push factors—particularly satisfaction—exert the strongest influence on career commitment, as evidenced by the high coefficient ($0.4$) in the satisfaction–career commitment relationship.
Although we did not explicitly apply the PPM framework, our findings align in demonstrating a strong correlation between satisfaction and career commitment, with obsolescence emerging as another significant predictor.
However, a notable difference is that in Fu's study, commitment was positively influenced by career investment, whereas in our study, we observed a slight negative effect on commitment.

Other research results on IT workers also inform our work. Brooks hypothesized that professional identification, commitment, and job satisfaction correlated with abandonment intention; findings showed that professional commitment reduced abandonment intention, while professional identification had an indirect effect \cite{brooks2015identifying}. Scholtz's study of IT professionals in South Africa identified significant correlations between career abandonment intention and factors such as job satisfaction, job insecurity, and adaptability to change \cite{scholtz-2019-role}. Armstrong and Bellini \cite{armstrong-2015-exhaustion,bellini-2019-should} investigated whether professional commitment, work exhaustion, job insecurity, and job satisfaction predicted abandonment intention. Differences in these factors' effects on software developers, compared to broader IT roles, remain an area for future research.

\section{Conclusion}
\label{conclusion}

In this paper, we surveyed 221 software developers to apply an adapted version of the Investment Model to explain career abandonment. Our contributions include:

\begin{itemize}
    \item Development of a theoretical model specifically focused on software developers' career abandonment intention, incorporating satisfaction factors intrinsic to the profession;
    \item Analysis of the relationship between work-related factors —- particularly satisfaction with core software development activities —- and the intention to leave a career in software development;
    \item Empirical support for the Investment Model's relevance in explaining career abandonment intention among software developers, highlighting a strong mediating effect of career commitment on the relationships among satisfaction, investment, alternatives, and career abandonment intention.
\end{itemize}

These results bring further questions and opportunities for future research. 
First, it is essential, 
to account for cultural differences, 
to replicate this study in other regions or countries. 
Second, the model was adapted from the Investment Model, 
but alternative and modified versions of the theoretical model were not tested; 
model refinement is recommended in SEM \cite{hair-2009-analise}. 
Testing variations of the original model could further confirm its adequacy. 
In addition, 
other models for career abandonment or career change could provide complementary insights into our results. 
For example, Social Cognitive Career Theory (SCCT) \cite{brown2023scct} emphasizes the role of self-efficacy, outcome expectations, and personal goals, 
which might intersect with the investment model's constructs of satisfaction and alternatives. 
We encourage future research to explore how these models interact and whether combining their perspectives can yield a more comprehensive understanding of the factors driving career abandonment in software development.


\section*{Acknowledgment}
\label{ack}

The authors thank Narallynne Araújo, Francielle Silva and Camila Sarmento for their research support. 
OpenAI's ChatGPT assisted with statistical analysis 
(a few R functions were generated from a high-level description provided by the authors)
and english grammar and style refinement 
(out of text written by the authors, provided as input).
Google's NotebookLM was used for summarizing ideas and contributions from the related work.
No sections or paragraphs were written completely from scratch by a GenAI tool.
This work was partially supported by CAPES, by providing scholarship for one of the authors. 


\appendix
\label{append}

\small

The full survey instrument is listed below (except for demographic questions).

\subsection*{Documentation}
\begin{itemize}
    \item \textbf{DOC1}: The documentation produced in my current job is below expectations.
    \item \textbf{DOC2}: My current job requires me to work with insufficient documentation.
    \item \textbf{DOC3}: In my current job project documentation is deprioritized.
    \item \textbf{DOC4}: My current job requires a low documentation load.
\end{itemize}

\subsection*{Requirements}
\begin{itemize}
    \item \textbf{REQ1}: In my current job requirements are not well defined.
    \item \textbf{REQ2}: The quality of the requirements affects how well my task is performed.
    \item \textbf{REQ3}: In my current job, the number of requirement changes is higher than expected.
    \item \textbf{REQ4}: In my current job, the requirements are not as clear as they should be.
\end{itemize}

\subsection*{Rework}
\begin{itemize}
    \item \textbf{REW1}: In my current job, I often have to redo an activity more than once.
    \item \textbf{REW2}: In my current job, I often redo a feature due to a planned or unplanned project change.
    \item \textbf{REW3}: In my current job, I often redo a task because the expected outcome wasn't achieved.
    \item \textbf{REW4}: In my current job, I often have to redo an activity, partially or fully, due to human error or guideline changes.
\end{itemize}

\subsection*{Planning}
\begin{itemize}
    \item \textbf{PLAN1}: In my current job, estimatives are often imprecise.
    \item \textbf{PLAN2}: In my current job, the deadline affects how well the task is performed.
    \item \textbf{PLAN3}: In my current job, I often have more tasks than expected.
    \item \textbf{PLAN4}: In my current job, I am often assigned tasks outside my area of expertise.
\end{itemize}

\subsection*{Code Quality}
\begin{itemize}
    \item \textbf{CODE1}: In my current job, the code quality of a project is below expectations.
    \item \textbf{CODE2}: In my current job, the code quality of a project affects the quality of the deliverables.
    \item \textbf{CODE3}: In my current job, the number of metrics and standards in a project affects task delivery.
    \item \textbf{CODE4}: In my current job, code readability is below expectations.
\end{itemize}

\subsection*{Threat of Professional Obsolescence}
\begin{itemize}
    \item \textbf{OBS1}: The technical knowledge required for my current job changes constantly.
    \item \textbf{OBS2}: Some technologies that were once valuable and important in my professional field are no longer useful.
    \item \textbf{OBS3}: The technical knowledge in my profession becomes outdated quickly.
    \item \textbf{OBS4}: My professional knowledge becomes obsolete at a fast pace.
\end{itemize}

\subsection*{Availability of Career Alternatives}
\begin{itemize}
    \item \textbf{ALT1}: If I want, I can easily get a comparable or better career opportunity than my current one.
    \item \textbf{ALT2}: There are enough career options for me to work in the job market.
    \item \textbf{ALT3}: It will be difficult to switch to another career.
\end{itemize}

\subsection*{Career Investment}
\begin{itemize}
    \item \textbf{INV1}: I have invested a lot in my professional career.
    \item \textbf{INV2}: I have invested in specific training for my professional career.
    \item \textbf{INV3}: I have invested time working in my professional career.
\end{itemize}

\subsection*{Career Commitment}
\begin{itemize}
    \item \textbf{CMT1}: The discomforts associated with the software development career sometimes seem too great.
    \item \textbf{CMT2}: Given the problems I encounter in my software development career, I sometimes wonder if I can get enough out of it.
    \item \textbf{CMT3}: Given the problems in the software development career, I sometimes wonder if the personal burden is worth it.
    \item \textbf{CMT4}: I strongly identify with the idea of a career in software development.
\end{itemize}

\subsection*{Intention to Abandon}
\begin{itemize}
    \item \textbf{ABDN1}: I intend to continue working in the software development career until I retire.
    \item \textbf{ABDN2}: I expect to work in a career different from software development at some point in the future.
    \item \textbf{ABDN3}: I often think about leaving the software development career.
    \item \textbf{ABDN4}: Likely, I will soon explore career opportunities outside the software development field.
\end{itemize}

\bibliographystyle{IEEEtran}
\bibliography{aband}

\end{document}